# On the Quantization Procedure in Classical Mechanics and the Reality of Bohm's Ψ-Field

*V. D. Rusov*

*Odessa National Polytechnic University*, *Odessa, Ukraine, siiis@te.net.ua*
*Bielefeld University*, *Bielefeld, Germany*

Using Chetaev's theorem on stable trajectories in dynamics in the presence of perturbation forces we obtain a generalized stability condition for Hamiltonian systems that has the form of the Schrödinger equation. We show that the energy of the perturbation forces generating generalized Chetaev's stability condition is of electromagnetic nature and exactly coincides with Bohm's "quantum" potential.

We stress that not taking into account the reality of Bohm's electromagnetic $\psi$–field turns out to be the fundamental reason of violation of the famous Bell inequality, in the recent most precise direct experiments, as well as in theoretical calculations based on the simple Wigner model. We discuss the possibility of direct experiments, alternative to Bell's ideology and related, e.g. with observation of nuclear stochastic resonance during the $\alpha$–decay process, which cannot occur in the framework of the probabilistic interpretation of quantum mechanics, but has to occur in the alternative Bohm interpretation.

In this Letter we analyse the question, which can be formulated the following rather strict and paradoxical form: "Are the so-called quantization conditions that are imposed on the corresponding spectrum of a dynamical system possible in principle, analogously to what is taking place in quantum mechanics?"

Surprisingly, the answer to this question is positive and has been given more than 70 years ago by the Russian mathematician N. G. Chetaev in his article «On stable trajectories of dynamics» [1,2]. The leading idea of his work and of the whole scientific ideology was the most profound personal paradigm, with which begins, by the way, his principal work [2]: «Stability, which is a fundamentally general phenomenon, has to appear somehow in the main laws of nature». Here, it seems, Chetaev states for the first time the thesis of the fundamental importance of theoretically stable motions and of their relation to the motions actually taking place in mechanics. He explains it as follows: the Hamiltonian theory of holonomic mechanical systems being under the action of forces admitting the force function has well proven itself, although, as Liapunov has shown, arbitrarily small perturbation forces can theoretically make such stable motions unstable. Nevertheless, in nature precisely and only stable motions preserve the general characteristics of motion, and thus, only they more or less correctly describe the motions actually taking place. Finally, Chetaev come to a conclusion, which at first sight may seem paradoxical that these arbitrarily small perturbation forces or, more precisely, "small dissipative forces with full dissipation, which always exist in our nature, represent a guaranteeing force barrier which makes negligible the influence of nonlinear perturbation forces" [3]. Furthermore, it has turned out that this "clear stability principle of actual motions, which has splendidly proven itself in many principal problems of celestial mechanics… unexpectedly gives us a picture of almost quantum phenomena" [4]. A similar point of view can be found in Dirac [5] where a quantization procedure appears in the framework of generalized Hamiltonian dynamics which is connected with the selection of the so-called small integrable *A*-spaces, only in which solutions of the equations of motion, and thus, only stable motions of a physical system are possible (see Eqs. (48) in Ref. [5]).

Below we generalize the Chetaev's theorem [1,2] in case when the Hamiltonian of system is time-dependent. For that let us consider a material system (where $q_1,\ldots, q_n$ and $p_1,\ldots, p_n$ are generalized coordinates and momenta of a holonomic system) in the field of potential forces admitting the force function of $U(q_1,\ldots, q_n)$ type.

In the general case, when the action $S$ is an explicit function of time, the complete intergal of the Hamilton-Jacobi differential equation corresponding to the system under consideration has the form

$$S = f(t, q_1, ..., q_n; \alpha_1, ..., \alpha_n) + A, \qquad (1)$$

where $\alpha_1, ..., \alpha_n$ and A are arbitrary constants, and the general solution of the mechanical problem, according to the well-known Jacobi theorem is defined by the formulas

$$\beta_i = \frac{\partial S}{\partial \alpha_i}, \quad p_i = \frac{\partial S}{\partial q_i}, \quad i = 1, ..., n, \qquad (2)$$

where $\beta_i$ are new constants of integration. Possible motions of the mechanical system are determined by different values of the constants $\alpha_i$ and $\beta_i$.

We will call the motion of the material system, of which the stability is going to be studied, non-perturbed motion. To begin with, let us study the stability of such a motion with respect to the variables $q_i$ under the perturbation only of the initial values of the variables (i.e. of the values of the constants $\alpha_i$ and $\beta_i$) in absence of perturbation forces.

If we denote by $\xi_j = \delta q_j = q_j - q_j(t)$ and $\eta_j = \delta p_j = p_j - p_j(t)$ the variations of the coordinates $q_j$ and the momenta $p_j$, and by $H(q_1, ..., q_n, p_1, ..., p_n)$ the Hamilton function, then it is easy to obtain for Hamilton's canonical equations of motion

$$\frac{dq_j}{dt} = \frac{\partial H}{\partial p_j}, \quad \frac{dp_j}{dt} = -\frac{\partial H}{\partial q_j} \qquad (3)$$

differential equations (in first approximation) in Poincaré's variations [6], which have the following form

$$\frac{d\xi_i}{dt} = \sum_j \frac{\partial^2 H}{\partial q_j \partial p_i} \xi_j + \sum_j \frac{\partial^2 H}{\partial p_j \partial p_i} \eta_j,$$

$$\frac{d\eta_i}{dt} = -\sum_j \frac{\partial^2 H}{\partial q_j \partial q_i} \xi_j - \sum_j \frac{\partial^2 H}{\partial p_j \partial q_i} \eta_j \qquad (i = 1, ..., n) \qquad (4)$$

where the coefficients are continuous and bounded real functions of $t$. These equations are of essential importance in studies of the stability of motion of conservative mechanical systems. Let us show this.

Poincaré has found in Ref.[6] that if $\xi_s, \eta_s$ and $\xi'_s, \eta'_s$ are any two particular solutions of the variational equations (4), then the following quantity is invariant

$$\sum_s (\xi_s \eta'_s - \eta_s \xi'_s) = C, \qquad (5)$$

where C is a constant. The proof is just a differentiation over $t$.

It is not difficult to show that for each $\xi_s, \eta_s$ there is always at least one solution $\xi'_s, \eta'_s$ for which the constant C in Poincaré's invariant does not vanish. Indeed, for a non-trivial solution $\xi_s, \eta_s$ one of the initial values $\xi_{s0}, \eta_{s0}$ at time $t_0$ will be different from zero. Then the second particular solution can always be defined by the initial values $\xi'_{s0}, \eta'_{s0}$ in such a way that the constant under consideration does not vanish.

Let for two solutions of the variation equations $\xi_s$, $\eta_s$ and $\xi'_s, \eta'_s$ the value of the constant $C$ be different from zero, and $\lambda$ and $\lambda'$ are the characteristic functions corresponding to these solutions. If we apply to this invariant Liapunov's theory [7], then we can directly, on the one hand, conclude that the characteristic value of the left-hand-side of the invariance relation (5), corresponding to the non-vanishing constant, is zero. On the other hand, this allows us to obtain the following inequality:

$$\lambda + \lambda' \leq 0. \tag{6}$$

If we now assume that the system of Pincaré's variation equations is correct, "which is natural…, when we are dealing with nature…" [4], then using Liapunov's theorem on the stability of the systems of differential equations in first approximation [7], it is easy to show that for the stability of the non-perturbed motion of the Hamiltonian system under consideration it is necessary that all the characteristic numbers of the independent solutions in Eq.(7) be equal to zero:

$$\lambda = \lambda' = 0. \tag{7}$$

Thus, Eq (7) represents a stability condition for the motion of the Hamiltonian system (4) with respect to the variables $q_i$ and $p_i$ under the perturbation of the initial values of the variables only, i.e. the values of the constants $\alpha_i$ and $\beta_i$. However, the determination of the characteristic numbers as functions of $\alpha_i$ и $\beta_i$ is a very difficult problem and therefore is not practical. The problem becomes simpler, if we note that, since the non-perturbed motion of our Hamiltonian system satisfying condition (7) is stable under any perturbations of initial conditions, it has to be stable under arbitrary perturbations of the constants $\beta_i$ only. In other words, the problem is reduced to the determination of the so-called conditional stability.

According to this assumption about the character of initial perturbations from the solutions of the Hamilton-Jacobi equation (2) the following relations are obtained immediately up to the terms of the second order

$$\eta_i = \sum_j \frac{\partial^2 S_0}{\partial q_i \partial p_j} \xi_i, \tag{8}$$

which allows us, taking into consideration the relation

$$H = \frac{1}{2} \sum g_{ij} p_i p_j + U, \tag{9}$$

to write the first group of Eqs. (4) in the form

$$\frac{d\xi_i}{dt} = \sum_{js} \xi_s \frac{\partial}{\partial q_s} \left( g_{ij} \frac{\partial S_0}{\partial q_j} \right). \tag{10}$$

where coefficients $g_{ij}$ depend on coordinates only. Here the variables $q_j$ and the constants $\alpha_j$ on the right-hand sides must be replaced using their values corresponding to non-perturbed motion.

If the variation equations (10) are correct, then, according to the well-known Liapunov theorem [7] about the sum of the eigenvalues of the independent solutions and to condition (9), we can conclude that the following condition is necessary for stability (10):

$$\lambda\{\exp\int Ldt\}=0, \quad where \quad L=\sum_{ij}\frac{\partial}{\partial q_i}\left(g_{ij}\frac{\partial S_0}{\partial q_j}\right), \tag{11}$$

where $\lambda$ is the eigenvalue of the function in the brackets.

Furthermore, if the system of equations (10), in additions to the correctness giving condition (11), satisfies reducibility requirements and if the corresponding linear transformation

$$x_i = \sum_j \gamma_{ij}\xi_j \tag{12}$$

has a constant determinant $\|\gamma_{ij}\| \neq 0$, then, due to the invariance of the eigenvalues of the solutions of system (10) under such a transformation and due to the well-known Ostrogradsky-Liouville theorem, it can be shown that in this case from Eq.(11) a necessary stability will follow in the form

$$L = \sum_{ij}\frac{\partial}{\partial q_i}\left(g_{ij}\frac{\partial S_0}{\partial q_j}\right) = 0, \tag{14}$$

which expresses the vanishing of the sum of the eigenvalues of system (11). A simple but elegant proof of condition (13) can be found in Ref. [8].

Let us now consider a more complicated problem. Let the material system in actual motion is under the action of forces with force function $U$, presumably taken into account by the considerations above, and some unknown perturbation forces, which are supposed to be potential and admitting the force function $Q$. Then the actual motion of the material system will take place under the influence of the forces with the joint force function $U^* = U + Q$, and thereby the actual motion of the system will not coincide with the theoretically predicted (in the absence of perturbation).

Keeping the problem setting the same as above, which concerns the stability of the actual unperturbed motions under the perturbation of initial conditions only, the necessary condition for stability in first approximation of the type (13) will not be effective in the general case, since the new function $S$ is unknown (just as $Q$). Nevertheless it turns out that it is possible to find such stability conditions, which do not depend explicitly on the unknown action function $S$ and potential $Q$.

Thus, let us start with the stability requirement of the type (13), assuming that the conditions for its existence (well-definiteness etc.) for actual motions are fulfilled. Let us introduce in Eq. (13) instead of function $S$ a new function $\psi$, defined by

$$\psi = A\exp(ikS), \tag{14}$$

where $k$ is a constant and $A$ is a real function on the coordinates $q_i$ only.

From this follows

$$\frac{\partial S}{\partial q_j} = \frac{1}{ik}\left(\frac{1}{\psi}\frac{\partial \psi}{\partial q_j} - \frac{1}{A}\frac{\partial A}{\partial q_j}\right) \tag{15}$$

and, therefore, Eq. (13) will become

$$\sum_{i,j}\frac{\partial}{\partial q_i}\left[g_{ij}\left(\frac{1}{\psi}\frac{\partial \psi}{\partial q_j}-\frac{1}{A}\frac{\partial A}{\partial q_j}\right)\right]=0. \qquad (16)$$

On the other hand, for the perturbed motion it is possible to write the Hamilton-Jacobi equation in the general case when Hamiltonian H is time-dependent:

$$\frac{1}{2k^2}\sum_{i,j}g_{ij}\left(\frac{1}{\psi}\frac{\partial \psi}{\partial q_i}-\frac{1}{A}\frac{\partial A}{\partial q_i}\right)\left(\frac{1}{\psi}\frac{\partial \psi}{\partial q_j}-\frac{1}{A}\frac{\partial A}{\partial q_j}\right)=\frac{\partial S}{\partial t}+U_0+Q, \qquad (17)$$

where $\partial S/\partial t$ is obtained by Eq.(14). Adding Eqs. (17) and (18) we have a necessary stability condition (in the first approximation) in this form

$$\frac{1}{2k^2\psi}\sum_{i,j}\frac{\partial}{\partial q_i}\left(g_{ij}\frac{\partial \psi}{\partial q_j}\right)-\frac{1}{2k^2 A}\sum_{i,j}\frac{\partial}{\partial q_i}\left(g_{ij}\frac{\partial A}{\partial q_j}\right)-$$

$$-\frac{1}{k^2 A}\sum_{i,j}g_{ij}\frac{\partial A}{\partial q_j}\left(\frac{1}{\psi}\frac{\partial \psi}{\partial q_i}-\frac{1}{A}\frac{\partial A}{\partial q_i}\right)-\frac{1}{ikA\psi}\left[A\frac{\partial \psi}{\partial t}-\psi\frac{\partial A}{\partial t}\right]-U-Q=0. \qquad (18)$$

Equality (18) will not contain $Q$, if the amplitude $A$ is defined from the equation

$$\frac{1}{2k^2 A}\sum_{i,j}\frac{\partial}{\partial q_i}\left(g_{ij}\frac{\partial A}{\partial q_j}\right)+\frac{i}{kA}\sum_{i,j}g_{ij}\frac{\partial A}{\partial q_j}\frac{\partial S}{\partial q_i}-\frac{1}{ikA}\frac{\partial A}{\partial t}+Q=0, \qquad (19)$$

which, after the separation into the real and imaginary parts, splits into two equations

$$Q=-\frac{1}{2k^2 A}\sum_{i,j}\frac{\partial}{\partial q_i}\left(g_{ij}\frac{\partial A}{\partial q_j}\right), \quad \frac{\partial A}{\partial t}=-\sum_{i,j}g_{ij}\frac{\partial A}{\partial q_j}\frac{\partial S}{\partial q_i}. \qquad (20)$$

Here evident consequences from the expressions for $Q$ (20) arise. For a strict fulfilment of the inequality $Q\neq 0$, e.g. in case that $\partial A/\partial t=0$, it is necessary that, firstly, the direction of the disturbance wave always was normal to the particle velocity vector, and, secondly, some small gyroscopic forces constantly act on the disturbance wave and the particle. Physical content and explanation of the both conditions will be given later.

Thus, if the perturbing forces have the structure of the type (20) satisfying the requirement formulated above, the necessary stability condition (18) will have the form

$$\frac{i}{k}\frac{\partial \psi}{\partial t}=-\frac{1}{2k^2}\sum_{i,j}\frac{\partial}{\partial q_i}\left(g_{ij}\frac{\partial \psi}{\partial q_j}\right)+U\psi. \qquad (21)$$

An obvious analysis of Eq. (21) allows us to make the following conclusion. One-valued, finite and continuous solutions of Eq. (21) for the function $\psi$ in stationary case are admissible only for the eigenvalues of the total energy $E$ and, consequently, the stability of actual motions considered here, can take place only for these values of total energy $E$.

Let us now come back to the problem of quantization and illustrate it on a very simple example. Let us consider a material point of mass $m$ moving in the field of conservative forces with force function $U$, which in the general case is time-dependent. We will study stability of motion of this point in Cartesian coordinates $x_1$, $x_2$, $x_3$. Denoting the momenta coordinatewise as $p_1$, $p_2$, $p_3$ we obtain for the kinetic energy the well-known expression

$$T = \frac{1}{2m}(p_1^2 + p_2^2 + p_3^2). \tag{22}$$

In this case conditions (21) for the structure of the perturbation forces admit the following relations

$$Q = -\frac{\hbar^2}{2m}\frac{\Delta A}{A}, \quad \frac{\partial A}{\partial t} = \sum \frac{\partial A}{\partial x_i} p_i, \quad k = \frac{1}{\hbar}, \tag{23}$$

and the differential equation (21), which is used for the determination of stable motions, becomes

$$i\hbar \frac{\partial \psi}{\partial t} = -\frac{\hbar^2}{2m}\Delta \psi + U\psi, \tag{24}$$

i.e. coincides with the well-known Schrödinger equation in quantum mechanics [9], which represents a relation constraining the choice of the constants of integration (of the total energy $E$ in stationary case) of the full Hamilton-Jacobi integral.

In the case it becomes interesting to consider the case connected with the inverse substitution of the wave function (14) in the Schrödinger equation (24), which generates an equivalent system of equations, known as the Bohm's system of equations [10], however, taking into account conditions (13) and (20):

$$\frac{\partial A}{\partial t} = -\frac{1}{2m}[A\Delta S + 2\nabla A \cdot \nabla S] = -\nabla A \cdot \frac{\nabla S}{m}, \tag{25}$$

$$\frac{\partial S}{\partial t} = -\left[\frac{(\nabla S)^2}{2m} + U_0 - \frac{\hbar^2}{2m}\frac{\Delta A}{A}\right], \tag{26}$$

It is important to note that the last term in Eq. (28), which in interpretation [10] is of a "quantum" potential of the so-called Bohm's $\psi$–field [10] exactly coincides with the perturbation potential $Q$ in (23). At the same time Eq. (25) is identical with condition for $\partial A/\partial t$ in Eq. (20).

If we make a substitution of $P=\psi\psi^*=A^2$ type, the Eqs. (25) and (26) can be rewritten as follows

$$\frac{\partial P}{\partial t} = -\nabla P \cdot \frac{\nabla S}{m}, \tag{27}$$

$$\frac{\partial S}{\partial t} + \frac{(\nabla S)^2}{2m} + U - \frac{\hbar^2}{4m}\left[\frac{\Delta P}{P} - \frac{1}{2}\frac{(\nabla P)^2}{P^2}\right] = 0. \tag{28}$$

Following Bohm [10], we note that "if we consider an ensemble of particle trajectories which are solutions of the equations of motion, then it is a well-known theorem of mechanics that if all of these trajectories are normal to any given surface of constant $S$, then they are normal to all surfaces of constant $S$, and $\nabla S(x)/m$ will be equal to the velocity vector, $\vec{v}$ (x), for any particle passing the point $x$" [10].

The statement that $P(x,y,z)$ indeed is the probability density for a particle in our ensemble is substantiated as follows. Let us assume that the influence of the perturbation forces generated by the potential $Q$ on the wave packet in an arbitrary point in the phase space is proportional to the density of the trajectories ($\psi\psi^*=A^2$) at this point. From where follows that the wave packet is practically not perturbed when the following condition is fulfilled

$$\int Q\psi\psi^* dV \Rightarrow \min, \quad where \quad \int \psi\psi^* dV = 1, \tag{29}$$

where $dV$ denotes a volume element of the phase space. And this implies in turn that, for the totality of motions in the phase space, the perturbation forces allow the absolute stability only if condition (29) is fulfilled or, in other words, when the obvious condition for the following variational problem is fulfilled:

$$\delta Q = Q - Q_\varepsilon = 0. \tag{30}$$

This variational principle (30) is actually nothing else than the principle of least action of perturbation of Chetaev [3].

Let us write for $Q$ (using the previous notations and Eq. (9)) the following equality

$$Q = -\frac{\partial S}{\partial t} - U - T = -\frac{\partial S}{\partial t} - U - \frac{1}{2}\sum_{ij} g_{ij} \frac{\partial S}{\partial q_i} \frac{\partial S}{\partial q_j}. \tag{31}$$

On the other hand, if (14) holds, it is easy to show that

$$\frac{1}{2}\sum_{ij} g_{ij} \frac{\partial S}{\partial q_i} \frac{\partial S}{\partial q_j} = -\frac{1}{2k^2\psi^2}\sum_{ij} g_{ij} \frac{\partial \psi}{\partial q_i} \frac{\partial \psi}{\partial q_j} + \frac{1}{2k^2 A^2}\sum_{ij} g_{ij} \frac{\partial A}{\partial q_i} \frac{\partial A}{\partial q_j} +$$

$$+ik\frac{1}{2k^2 A^2}\sum_{ij} g_{ij} \frac{\partial A}{\partial q_i} \frac{\partial S}{\partial q_j}. \tag{32}$$

Further, we need to carry out the following successive substitutions. First, for the first term on the right-hand side of Eq.(32) we substitute its value from the first stability condition (16), then we insert the obtained relation into (32) and finally put the result into the equation corresponding to the variational principle (31).

It is remarkable that as a consequence of the substitution procedure described above we obtain a relation which exactly equal to Eq.(18) and, therefore, the resulting structure expression and the necessary condition for stability coincide with (20) and (21).

And this means that based on Chetaev's variational principle we obtained an independent confirmation of the fact that the mathematical nature of $P(x,y,z)$ indeed reflects not simply the notion of probability density of "something" according to Bohm's equation of continuity [10], but plays the role appropriate to the probability density of the number of particle trajectories.

Such physical content of the probability density function $P(x,y,z)$ and simultaneously, the exact coincidence of the "quantum" potential of Bohm's $\psi$ field [10] in equation (26) and of the force function of perturbation $Q$ in Eq.(23) leads to astonishing, but fundamental conclusions:

– the entire laborious, but brilliant experience of the traditional probabilistic quantum mechanics not only confirms directly the effectiveness of Schrödinger's equation as instrument for description of the physical reality, but also confirms the relevance of the assumptions needed for the derivation of the necessary stability condition (21);

– in view of Chetaev's theorem about the stable trajectories of dynamics the reality of Bohm's $\psi$–field is an evident and indisputable fact, which in turn leads to the at first glance paradoxical conclusion that classical and quantum mechanics are two complementary procedures of one Hamiltonian theory. In the framework of this theory Eq. (26) is an ordinary Hamilton-Jacobi equation and differs from the analogous equation obtained from $\hbar \to 0$ ($Q=0$) only in so far as its solution is a priori stable. In other words, classical mechanics and the quantization (stability) conditions represent, in contradiction to the correspondence principle, two complementary procedures for description of stable motions of a physical system in a potential field. This is characteristic for the problems of the classical atomic theory [11] and becomes apparent with a perplexing precision in the theoretical treatment of the well-known collision experiments in the framework of classical mechanics [12-16].

Analysis of Eq. (24) naturally poses an extremely deep and fundamental question about the physical nature of really existent (as is shown above) small perturbing forces, or "small dissipative forces with total dissipation" according to Chetaev [2].

We consider that we deal with perturbation waves of de Broglie type, whose action describes by Bohm $\Psi$-field. Such conclusion is caused, first of all, by the fact that de Broglie's "an embryonic theory of waves and particles union" [17] was developed just on the basis of identity of least action principle and the Fermat principle, that very exactly and clearly reflects a physical essence of the Chetaev's theorem on the stable trajectories of dynamics (see Eq.(30)).

Such assumption can be views not only as unacceptable, but also unpardonably erroneous if it were not the results of the Polish physicist Gryzinski. He has shown that, using real de Broglie waves in the form of oscillating electromagnetic field of photon or electron caused by translational precession of the spin, it is possible to explain the particle interference and diffraction phenomena [11,18] in the framework of Newtonian mechanics and classical electrodynamics. He also has shown, using the concept of localized electron, i.e. in the framework of classical dynamics, how electron spin axis precession hides behind the Sommerfeld quantization integral, and how alternating electromagnetic field caused by the precession of its magnetic axis hides behind the wave field of the electron [11,18].

In this sense it is interestingly to consider notion of translation precession of the particle spin from the standpoint of the Chetaev's theorem of the dynamics stable trajectories. It is not difficult to see that at $\partial A/\partial t=0$ the translational precession of the spin around particle magnetic axis is just such gyroscopic force, which automatically provide satisfaction of special requirements to the structure of items in Eq. (20) and, thereby, ensures the particle stable motion. In other words, the Gryzinski assumption of the translational precession of the spin generating de Broglie electromagnetic disturbing wave follows naturally at $\partial A/\partial t=0$ from conditions (20) of the generalized Chetaev's theorem on the stable trajectories of dynamics.

It is obvious, that if this concept of de Broglie waves is correct, the next step should be a construction of algoristic theory of atomic collusions based on the solution of classical two-body problem, which makes it possible to trace the electron motion within the atom. It is astonishing, but this problem was actually solved by Gryzinski [12], where the complete set of relations strictly describing the two-body collision problem was obtained. The verification of this theory by famous experiments of Helbig and Everhardt [19] on electron capture by the proton at head-on collision with atomic hydrogen as well as by some other collision experiments of electrons, protons and hydrogen atoms [20] made it possible not only correctly and exactly to describe this experiments, but to obtain also the proof of non-trivial electron kinetics within the atom, namely radial (!) kinetics of electron in the nuclear Coulomb field [11,15,21]. Further applications of free-fall atomic model and of the results of classical collision theory allowed to precisely describe a number of important problems using the classical mechanics motion equations and Coulomb law, for example, ionization by collision and the excitation of molecules and atoms [12,13], Ramsauer effect [14] and van der Waals forces [15], atomic diamagnetism [16], atomic energy levels shift [18,22] and many others [18].

In fact, the number of problems, which were solved using the classical dynamics and notion of small perturbation (in other words, using the physically separated de Broglie wave, which generate by the spin), has reached the point when it is impossible to ignore the philosophical and physical aspects of this explicitly non-random circumstance.

Cause of practically complete ignoring of alternative theories satisfying Einstein's locality principle [9], apparently, is directly connected with the reliably established fact of Bell's inequality violation in the numerous experiments, the results of which, as is customary to consider, are evidence of the fact that quantum-mechanical predictions are not compatible with Bell's inequality. However, as it follows from our results, the unconscious ignoring the really existing electromagnetic Bohm's $\Psi$-field is the principle cause of the known Bell's inequality violation in the modern super-precision direct experiments as well as in the theoretical computations on the basis of the simple Wigner's model [9]. In other words, it is for this reason that real particle which has spin, for example, photon or electron, will not be described by the pure state because it will always be wrapped by the real field "fur coat" in the form of de Broglie electromagnetic wave, which is responsible for nonlocality state, the Bell's inequality violations will always exist in such experiments.

Because of above mentioned difficulties the experimental verification of Bell's inequality becomes practically impracticable. Therefore the new experiments alternative to the Bell's ideology should be made on another principle, for example, using direct determination of actually wave function by the ultrasensitive detection of electromagnetic perturbation wave interference (i.e., $\Psi$ trajectories), which accompanies electron diffraction, using the low intensity source as in Tonomura experiment [23]. At the same time the positive result can simultaneously become a basis for clear classic substantiation of the Aharonov-Bohm effect [24]. A different way consist in the direct proof of existence or, vice versa, complete absence of particles tunneling, which is possible within the framework of probabilistic quantum mechanics and is absolutely impossible within the framework of alternative theories satisfying Einstein's locality principle [9, 10]. For example, within the framework of Bom's causal interpretation [10] $\alpha$-decay can be described by the classic way using well-known Kramers' diffusion mechanism [25]. Obviously, that introduce of the external periodic signal into Kramers' equation (of Langevin type) must result in the observation of stochastic resonance [26]. In other words, the experiments on the search of nuclear stochastic resonance during $\alpha$-decay, which can not in principle take place within the framework of probabilistic interpretation of quantum mechanics but must be observed within the framework of the Bohm's interpretation, can be a worth alternative to the Bell's ideology.

At last, note that when first researcher UV-lasers of frequency about $10^{18}$–$10^{20}$ $s^{-1}$ [27] will appear in the near future, the problem of the stochastic mode excitation in the atomic nucleus under action of periodic external field will become actual not only in respect to the direct study of self-organization and quantum chaos in it, but fundamental in the view of realization of "tunnel" experiments alternative to the Bell's type experiments.